\documentclass[letterpaper, 10 pt, conference]{ieeeconf}
\IEEEoverridecommandlockouts  
\let\labelindent\relax
\usepackage{enumitem}

\usepackage{mathtools,amsthm}

\usepackage{url}
\usepackage{hyperref}
\usepackage{graphics} 
\usepackage{epsfig}
\usepackage{times} 
\usepackage{amsmath,amssymb,amsfonts}
\usepackage{algorithmic}
\usepackage{graphicx}
\usepackage{textcomp}
\usepackage{varioref}
\usepackage{import}
\usepackage{framed,xcolor}
\usepackage{color}
\usepackage{comment}
\usepackage{epstopdf}   
\usepackage{psfrag}
\usepackage{breqn}
\usepackage{float}
\usepackage{booktabs}
\usepackage{soul}
\usepackage{amsbsy}
\usepackage[bottom]{footmisc}
\usepackage{lineno}
\usepackage{cleveref}
\usepackage{microtype}
\usepackage{comment}
\usepackage{arydshln}
\usepackage{mathtools}
\newtheorem{Theorem}{Theorem}

\newtheorem{Lemma}{Lemma}

\newtheorem{Remark}{Remark}

\newtheorem{Assumption}{Assumption}

\usepackage{color}

\newcommand{\vect}[1]{\ensuremath{\boldsymbol{\mathrm{#1}}}}

\usepackage[normalem]{ulem} 
\usepackage[usenames,dvipsnames]{pstricks}
 \usepackage{epsfig}
 \usepackage{pst-grad} 
 \usepackage{pst-plot} 
 \usepackage[space]{grffile} 
 \usepackage{etoolbox} 

\def\r{\rho}
\def\f{\varphi}
\DeclareMathOperator{\G}{\Box}
\DeclareMathOperator{\F}{\rotatebox[origin=c]{45}{$\Box$}}

\allowdisplaybreaks

\title{\LARGE \bf
Distributionally Robust Control for  Chance-Constrained\\ Signal Temporal Logic Specifications
}

\author{Arash Bahari Kordabad, Eleftherios E.Vlahakis, Lars Lindemann, Dimos V. Dimarogonas, and Sadegh Soudjani
\thanks{Arash Bahari Kordabad and Sadegh Soudjani are with the Max Planck Institute for Software Systems, Kaiserslautern, Germany. E-mail: {\tt\small\{arashbk, sadegh\}@mpi-sws.org.} Eleftherios E.Vlahakis and  Dimos V. Dimarogonas are with the Division of Decision and Control Systems, KTH Royal Institute of Technology, Stockholm, Sweden. E-mail: {\tt\small\{vlahakis, dimos\}@kth.se}.  Lars Lindemann is with the Thomas Lord Department of Computer Science, Viterbi School of Engineering, University of Southern California, Los Angeles, USA. 
Email: {\tt\small llindema@usc.edu}.\newline This research is supported by the following grants: EIC 101070802 and ERC 101089047.}}

\begin{document}

\maketitle
\thispagestyle{empty}
\pagestyle{empty}

\begin{abstract}
We consider distributionally robust optimal control of stochastic linear systems under signal temporal logic (STL) chance constraints when the disturbance distribution is unknown. By assuming that the underlying predicate functions are Lipschitz continuous and the noise realizations are drawn from a distribution having a concentration of measure property, we first formulate the underlying chance-constrained control problem as stochastic programming with constraints on expectations and propose a solution using a distributionally robust approach based on the Wasserstein metric. We show that by choosing a proper Wasserstein radius, the original chance-constrained optimization can be satisfied with a user-defined confidence level. A numerical example illustrates the efficacy of the method.


\end{abstract}


 




\section{Introduction}


Control of stochastic systems under temporal logic finds application in a wide range of domains, including robotics, autonomous systems, and cyber-physical systems. The formal specification of system properties that can be formulated in a probabilistic setting, enabling a systematic approach to quantifying uncertainty and handling feasibility, lies at the core of this problem~\cite{Haesaert2018}. Signal temporal logic (STL) is a formal language that allows us to encode time-constrained tasks using both Boolean and quantitative semantics~\cite{donze2010robust, maler2004monitoring}. When systems are subject to stochastic disturbances and STL specification, a typical probabilistic approach is to formulate the problem as a chance-constrained program (CCP)~\cite{farahani2017shrinking}.
 



Most recent results in the probabilistic STL context focus on applying probability or risk measures to individual predicates. To address the impact of critical tail events when STL formulas are violated,~\cite{LindemannCDC2020} propose Risk Signal Temporal Logic, incorporating risk constraints over predicates while preserving Boolean and temporal operators. Authors in~\cite{Sadigh2016} introduce probabilistic signal temporal logic, allowing expression of uncertainty by incorporating random variables into predicates. Similarly, in~\cite{Sadigh2018}, chance-constrained temporal logic formulates chance constraints as predicates to model perception uncertainty for autonomous vehicles. Stochastic temporal logic introduced in~\cite{Li2017} is similar in syntax to chance-constrained temporal logic but is designed for stochastic systems, where perturbations affect system dynamics rather than predicate coefficients. 



Top-down approaches study STL probabilistic verification of stochastic systems considering chance constraints on the entire specification~\cite{Scher2022}. More closely related to our work is~\cite{FarahaniTAC2019}, in which the authors transform chance constraints into linear constraints using concentration of measure inequalities to provide a conservative approximation of the feasible domain. Due to the nonconvex feasible domains typically induced by CCPs, many studies focus on numerical methods to handle CCPs, such as {randomized optimization} where the original optimization is approximated by a scenario program (SP) by sampling the uncertainty space. SP approaches have been studied for convex~\cite{Campi2011,Calafiore2010}, and nonconvex CCPs~\cite{garatti2024non}.

Unlike stochastic optimization settings where the probability distribution is assumed to be known, distributionally robust optimization (DRO) addresses the lack of information on the probability distribution by considering the worst-case distribution within an ambiguity set. Various methods exist for constructing ambiguity sets, such as moment ambiguity~\cite{delage2010distributionally}, Kullback–Leibler divergence-based ball~\cite{hu2013kullback}, and Wasserstein-based ball~\cite{pflug2007ambiguity}. The Wasserstein ambiguity set represents a statistical ball within the space of probability distributions surrounding the empirical distribution, with its radius measured using Wasserstein distance. Wasserstein DRO offers a probabilistic guarantee based on finite samples within a tractable formulation~\cite{mohajerin2018data} and has attracted significant attention recently~\cite{kordabad2022safe}.



In the DRO literature, several works have addressed CCP directly. An explicit reformulation for both individual and joint CCPs was presented in~\cite{chen2024data}, where uncertainties are modeled as affine functions. The authors of~\cite{xie2021distributionally} reformulated CCP as a conditional value-at-risk (CVaR) mixed-integer program for affine functions. A more general approach for CCP has been proposed in~\cite{hota2019data}. However, dealing with expectation-constrained programs (ECPs) preserves linearity and convexity, making it often computationally simpler and more straightforward compared to evaluating or approximating CCPs, particularly for non-standard distributions.

An SP approach has been proposed in~\cite{Soudjani2018} for solving a general CCP by transforming it into an ECP assuming that 1) the underlying distribution of the uncertain parameters satisfies a \textit{concentration of measure} property and exhibits bounded variance, and 2) the constraint function is Lipschitz continuous in the uncertainty parameters.


In this paper, we formulate a stochastic optimal control problem as a CCP, where the underlying system is subject to stochastic disturbances with unknown distribution and a set of STL specifications. We first assume that the disturbance realizations follow a concentration of measure property and that the underlying predicate functions involved in the STL constraints are Lipschitz continuous in the perturbation parameters. We build upon the results in~\cite{Soudjani2018} and transform the CCP to stochastic programming with expectation constraints. Since the exact distribution is assumed to be unknown, we propose a data-driven Wasserstein distributionally robust approach that guarantees STL satisfaction in a probabilistic sense.

The remainder of the paper is organized as follows. In Section~\ref{sec:Prelim}, we present the system setup, the STL formulation and the control synthesis problem we study. The construction of a stochastic program, its connection to the original CCP through the concentration of measure, and the distributionally robust solution of the stochastic program are in Section~\ref{sec:Results}. A numerical study is in Section~\ref{sec:examples}, and concluding remarks are in Section \ref{sec:concl}.


\section{Problem Formulation}
\label{sec:Prelim}
\subsection{Discrete-Time Stochastic Linear Systems}
\label{sec:linsys}
We consider systems in discrete time with state space $\mathcal{X}\subseteq\mathbb{R}^{n}$, input space $\mathcal{U}\subseteq\mathbb{R}^{m}$, and disturbance set $\mathcal{W}\subseteq\mathbb{R}^n$, that can be modeled by linear  difference equations perturbed by stochastic disturbances:
\begin{equation}
\label{eq:discsys}
x_{k+1} = A x_k + B u_k + w_k,\;\;
\end{equation}
where $x_k\in\mathcal{X}$ denotes the state of the system at time instant $k$, $u_k\in\mathcal{U}$ denotes the control input at time instant $k$, and $w_k\in\mathcal{W}$ 
is a random vector that has an unknown probability distribution $\mathcal{P}$ supported on $\mathcal{W}$. Matrices $A\in\mathbb{R}^{n\times n}$, and $B\in\mathbb{R}^{n\times m}$, and initial state $x_0$ are assumed to be known. In the view of~\eqref{eq:discsys}, for any $k\in\mathbb N$, $x_k$ is a function of $x_0$, input sequence vector ${u}_{0:k}:=[u^\top_0,\ldots,u^\top_{k-1}]^\top$, 
and the process noise ${w}_{0:k}:=[w^\top_0,\ldots,w^\top_{k-1}]^\top$:
\begin{equation}
\label{eq:state}
x_{k} = A^{k} x_0 + \sum_{i=0}^{k-1}A^{k-i-1} \left(B u_i + w_i\right).
\end{equation}


\subsection{STL specifications}
\label{sec:stl}
We consider signal temporal logic (STL) formulas defined recursively according to the grammar~\cite{MalNic:04}: 
\begin{equation*}
\varphi ::= \mathcal{T}\mid \pi \mid \neg \varphi \mid\varphi \land \psi  \mid \varphi\, {U}_{[a,b]}\,\psi,
\end{equation*}
where $ \mathcal{T}$ is the \emph{true} predicate; $\pi$ is a predicate whose truth value is determined by the sign
of a predicate function of state variables, i.e. $\pi = \{\alpha(x)\ge 0\}$ with $\alpha:\mathbb R^n\rightarrow\mathbb R$;
$\psi$ is an STL formula;
$\neg$ and $\land$ indicate negation and conjunction of formulas;
and ${U}_{[a,b]}$ is the \emph{until} operator with $a,b\in\mathbb{N}$. A finite run $\xi := \{x_0,x_1,x_2,\dots, x_N\}$ satisfies $\varphi$ at time $k$, 
denoted by $(\xi,k) \models \varphi$ with the
Boolean semantics of STL formulas defined as follows:
  \begin{flalign*}
        & ( \xi,k)\models \pi \quad\quad\quad\, \Leftrightarrow \quad  \alpha( x_k)\geq 0,\\
        & ( \xi,k)\models \lnot \varphi \quad \quad\,\,\, \Leftrightarrow \quad  \lnot(( \xi,k)\models \varphi),
        \\
        & ( \xi,k)\models \varphi \land \psi \quad\,\,  \Leftrightarrow \quad  ( \xi,k)\models\varphi \wedge ( \xi,k)\models\psi, \\
        & ( \xi,k)\models \varphi\,{U}_{[a,b]}\, \psi  \Leftrightarrow  \,\,\,\, \exists k'\in\{a,\ldots, b\}, ( \xi,k+k')\models \psi  \\ &\,\,\,\qquad \qquad\qquad \qquad\quad \wedge\forall k''\in\{k, \ldots, k'\}, ( \xi,k'')\models\varphi,
    \end{flalign*}
Additionally, we derive the \emph{disjunction} operator as $\varphi \lor \psi:=\neg(\neg\varphi\land\neg\psi)$,
the \emph{eventually} operator as $\F_{[a,b]}\varphi := \mathcal{T}{ U}_{[a,b]} \varphi$,
and the \emph{always} operator as $\G_{[a,b]}\varphi:=\neg \F_{[a,b]}\neg\varphi$. Thus $(\xi,k) \models \F_{[a,b]} \varphi$ if $\varphi$ holds at some time instant between $a+k$ and $b+k$ and
$(\xi,k) \models \G_{[a,b]} \varphi$ if $\varphi$ holds
at every time instant between $a+k$ and $b+k$.

\smallskip\noindent\textbf{STL Robustness.}
In contrast to the above Boolean semantics, the quantitative semantics (a.k.a. robustness function) of STL~\cite{req_mining_hscc2013} assigns to each formula $\f$ a real-valued function $\r^\f$ of signal $\xi$ and $k$ such that $\r^\f> 0$ implies $(\xi,k) \models \f$. The robustness of a formula $\varphi$ with respect to a run $\xi$ at time $k$ is defined recursively as
\begin{flalign*}
  &\quad  \rho^{\top}(\xi,k)= +\infty &\\
        &\quad  \rho^{\mu}(\xi,k)= \alpha (x_k) &\\
        &\quad  \rho^{\lnot\phi}(\xi,k)= -\rho^{\phi}(\xi,k)&\\
        &\quad  \rho^{\phi \wedge \psi}(\xi,k)=\min(\rho^{\phi}(\xi,k),\rho^{\psi}(\xi,k))&\\     
&\quad  \rho^{\phi\, {U}_{[a,b]}\, \psi}(\xi,k)=\max_{k'\in\{a,\ldots,b\}}\Big(\min\big(\rho^{\psi}(\xi,k+k'),\\ &\qquad\qquad\qquad\qquad\qquad\qquad\quad\,\, \min _{k''\in\{k,\ldots,k'\}}\rho^{\phi}(\xi,k'')\big)\Big)\,.
\end{flalign*}
Therefore, the value of the robustness function $\rho^{\phi}(\xi,k)$ can be interpreted as how much the trajectory $\xi$ satisfies a given STL formula $\phi$. The robustness of the formulas $\F_{[a,b]}\varphi$ and $\G_{[a,b]}\varphi$ are
\begin{flalign*}
    \r^{\F_{[a,b]}\varphi}(\xi,k)= & \max_{k'\in\{a,\ldots, b\}} \r^{\varphi}(\xi,k+k'),\nonumber\\
    \r^{\G_{[a,b]} \varphi}(\xi,k) =&\min_{k'\in \{a,\ldots, b\}}\r^{\varphi}(\xi,k+k').\nonumber
\end{flalign*}
%
%
%

As described above, the robustness function is commonly defined with its arguments being the system trajectory and the time index. However, for the scope of this study, it is more comprehensible to explicitly define the robustness as a function of the control input and the disturbance. In essence, as in~\eqref{eq:state}, the system trajectory is determined by the initial state, input, and disturbance sequence. Therefore, we define the dynamic-dependent function $\varrho^{\varphi}$ as follows:
\begin{equation*}
     \varrho^{\varphi} (u_{k:N},w_{k:N},x_k,k) :=\rho^{\varphi}(\xi,k).
\end{equation*}
Moreover, at time $k=0$ for a given $x_0$, we eliminate $x_k$ and $k$ from the argument of $\varrho^{\varphi}$ and define
\begin{equation*}
     \varrho_0^{\varphi} (\vect u,\vect w) := \varrho^{\varphi} (\vect u,\vect w,x_0,0), 
\end{equation*} 
where $\vect u:=u_{0:N}$ and $\vect w:=w_{0:N}$.

\smallskip

In the following, we make a regularity assumption on the STL predicate functions. This assumption allows us to establish the Lipschitz continuity of robustness functions, define a well-defined chance constraint for the robustness function,  and utilize the concentration of measure property.

\begin{Assumption}\label{Assum:predicate}
    We assume that the predicate functions are Lipschitz functions.
\end{Assumption}

\subsection{Chance-constrained optimization}
In the following, we introduce the STL chance-constrained optimization. Our aim is to provide a control input $\vect u$ such that the function $ \varrho_0^{\varphi}$ becomes positive, or specifically lower bounded by a pre-defined positive robustness level $r_0$. Since this function is affected by the unknown disturbance $\vect w$, the purpose would be to satisfy the inequality $$
    \varrho_0^{\varphi} (\vect u,\vect w)\geq r_0$$ with probability at least a given threshold. 
    We denote the $N$-fold product of the probability measure $\mathcal{P}$ by $P:=\mathcal{P}^N$ supported on the Cartesian product space $\mathbb{W}:=\mathcal{W}^N$, and we denote $\mathbb U:=\mathcal{U}^N$. We then define the probability space $(\mathbb{W},\mathcal F, P)$ and the following chance-constrained program (CCP):
\begin{equation}
\label{eq:CCP}
\mathrm{CCP:}\begin{cases}
\min_{\vect u\in{\mathbb{U}}} \mathbb E_{P}\left[J(\vect u,\vect w)\right],\\
\mathrm{s.t.}\,\, P\left\{\varrho_0^{\varphi}(\vect u,\vect w)\geq r_0\right\}\ge 1-\varepsilon,
\end{cases}
\end{equation}
where $\varepsilon\in(0,1)$ is the constraint violation tolerance and $J$ is a lower semi-continuous cost function. The aim of this optimization is to find an optimal control sequence that minimizes the expected value of a performance function $J$ while ensuring that the STL constraint, with robustness level $r_0$, is satisfied with a probability of at least $1-\varepsilon$. Under Assumption~\ref{Assum:predicate} and for the linear system in~\eqref{eq:discsys}, where the mapping from the disturbances to the system trajectory is continuous, the above optimization is well-defined and attains a solution if it is feasible~\cite{Soudjani2018}.
This is due to the continuity of the robustness function under Assumption~\ref{Assum:predicate}, which arises from the continuity of functions containing $\min$ and $\max$ operators with continuous arguments. 





\subsection{Concentration of measure property}
In this paper, we make the following assumptions on the distribution of $\vect w$. 

\begin{Assumption}
	\label{ass:light-tailed}
	(Light-tailed distribution). The distribution of random variable $\vect w$ is a Light-tailed distribution. More specifically, there exists $a>1$ such that $C:=\mathbb{E}_P[\exp{\|\vect w\|^a}]< \infty$.
\end{Assumption}
Assumption~\ref{ass:light-tailed} holds for many distributions, e.g., multivariate normal distribution, exponential distribution, log-normal distribution, and all distributions with bounded support. 
\begin{Assumption}
	\label{ass:concentration}
	(Concentration of Measure). There exists a monotonically decreasing function $h:\mathbb R^{\ge 0}\rightarrow[0,1]$ such that
	\begin{equation}
	\label{eq:concentration}
	 P\left\{|f(\vect w)-\mathbb E\left[f(\vect w)\right]|\le t\right\}\ge 1-h(t),\quad\forall t\ge 0,
	\end{equation}
	holds for any Lipschitz continuous function $f:\mathbb{W}\rightarrow\mathbb R$ with Lipschitz constant $1$.
\end{Assumption}

We recall that a function $f:\mathbb{W}\rightarrow \mathbb{R}$ is a Lipschitz continuous function if there exists $L\geq 0$ such that for any two vectors $\vect w_1,\vect w_2\in\mathbb{W}$, $
    \frac{|f(\vect w_1)-f(\vect w_2)|}{d_{\mathbb{W}}(\vect w_1,\vect w_2)}\leq L$ holds,  where $d_{\mathbb{W}}$ denotes a metric on the set $\mathbb W$. Constant $L$ is referred to as the Lipschitz constant. Throughout this paper, we use the 2-norm on $\mathbb{W}$ to calculate the Lipschitz constants, given by $d_{\mathbb{W}}(\vect w_1,\vect w_2)=\sqrt{\left \langle \vect w_1-\vect w_2 , \vect w_1-\vect w_2\right \rangle}$. Note that, by employing alternative metrics, the general results of the paper remain valid, and changing the metric only affects the values of the Lipschitz constant.

Assumption~\ref{ass:light-tailed} will be utilized in deriving a data-driven solution to ~\eqref{eq:CCP} in Section~\ref{sec:Results}. We use Assumption~\ref{ass:concentration} to construct a stochastic program for finding a (possibly sub-optimal) solution of CCP~\eqref{eq:CCP}. Note that Assumption~\ref{ass:concentration} also holds for many distributions. Examples of different distributions with the concentration of measure property and corresponding $h$ functions can be found in~\cite{Soudjani2018}. For instance,  the standard multi-variate Gaussian distribution satisfies \eqref{eq:concentration} with $h(t) = \min\{2e^{-2t^2/\pi^2},1\}$~\cite{barvinok1997measure,pisier1999volume}. Note that if we substitute $h(\cdot)$ with another monotonically decreasing function $\bar h(\cdot)$, and $\bar h(\cdot)$ satisfies $\bar h(\cdot)\geq h(\cdot)$, then the inequality \eqref{eq:concentration} remains valid when using $\bar h(\cdot)$.


In the following, we provide a lemma and a theorem that enable us to use the concentration of measure property in the context of STL and robustness function.

\begin{Lemma}\label{lem:lip:minmax}
    For any two Lipschitz functions $f_1:X\rightarrow\mathbb R$ and $f_2:X\rightarrow\mathbb R$, $\max(f_1, f_2)$ and $\min(f_1, f_2)$ are Lipschitz functions with $L := \max(L_1, L_2)$, where $L_i$ is the Lipschitz constant of $f_i$, $i\in\{1,2\}$.
\end{Lemma}
\begin{proof}
    The proof is given in the appendix.
\end{proof}
Note that Lemma~\ref{lem:lip:minmax} can be extended for the cases where we have more than two functions inside the $\min$ or $\max$ operators. In this case, one can readily verify that $\min(f_1,\ldots,f_n)$ and $\max(f_1,\ldots,f_n)$ are Lipschitz functions with constant $\max\{L_1,\ldots, L_n\}$ when $f_i$ is a  Lipschitz function with constant $L_i$ for all $i\in\{1,\ldots,n\}$. Moreover, the result holds for the combination of $\min$ and $\max$ with any number of operators. For instance, $\min(f_1,\ldots,\max(f_j,\ldots,f_{l}),\ldots, f_n)$ with $1\leq j\leq l\leq n$, is a Lipschitz function with constant $\max\{L_1,\ldots, L_n\}$. The following theorem uses Lemma~\ref{lem:lip:minmax} and shows that the robustness of an STL specification for linear systems is a Lipschitz function. Moreover, it provides the corresponding Lipschitz constant. 
\begin{Theorem}\label{the:lip}
    For any STL specification $\varphi$ with Lipschitz atomic predicates,  $\varrho_0^{\varphi}(\vect u,\vect w)$ is Lipschitz continuous with respect to $\vect w$ for the system defined in~\eqref{eq:discsys}. The Lipschitz constant will be $L_{\varphi} = L_1L_2$, where $L_1$ is the maximum Lipschitz constant of atomic predicates appearing in $\varphi$. More specifically, consider the STL formula $\varphi$ consists of $\mathcal J$ subformula with atomic predicate $\alpha_j$, $j\in \{1,\ldots, \mathcal J\}$, then:
\begin{equation*}
    L_1:= \max_{j\in \{1,\ldots, \mathcal J\}} L_{\alpha_j}, 
\end{equation*}
where $L_{\alpha_j}$ is the Lipschitz constant of $\alpha_j$ for $j\in\{1,\ldots, \mathcal J\}$. Constant $L_2$ is the maximum Lipschitz constant of $x_k$, the state at time $k\in\{1,2,\ldots,N\}$, with respect to $\vect w$, which is bounded by:
    \begin{equation}\label{eq:lip:the}
        L_2 = \sqrt{\sum_{i=0}^{N-1}\|A^i
        \|^2},
    \end{equation}
    where $\|A^i
        \|$ is the induced 2-norm of the matrix $A^i$ and $N$ is the length of the sequence $\vect w$.
    
\end{Theorem}
\begin{proof}
    The proof is given in the appendix.
\end{proof}

Theorem~\ref{the:lip} enables us to use the results of~\cite{Soudjani2018} in the context of STL by providing an explicit formula for the Lipschitz constant.
The next section details the proposed solution for the CCP in~\eqref{eq:CCP}.
\section{Solution Approach}\label{sec:Results}

In the following, we use Assumption~\ref{ass:concentration} for the Lipschitz continuous function $\varrho_0^{\varphi}(\vect u,\vect w)$ (cf. Theorem~\ref{the:lip}) and provide an under approximation for CCP~\eqref{eq:CCP}.
\begin{Theorem}\label{the:EP} Under Assumption~\ref{ass:concentration}, the feasible domain of the CCP~\eqref{eq:CCP} includes the feasible domain of the following expectation-constrained program (ECP):
 \begin{equation}
\label{eq:EP}
\mathrm{ECP:}\begin{cases}
\min_{\vect u\in\mathbb U} \mathbb E_{P} \left[J(\vect u,\vect w)\right],\\
\mathrm{s.t.}\,\,\mathbb E_{P} \left[\varrho_0^{\varphi}(\vect u,\vect w)\right]-L_\varphi h^{-1}(\varepsilon)\geq r_0,
\end{cases}
\end{equation}
    where $h$ is given by the class of distribution and $L_\varphi$ is the Lipschitz constant of $\varrho_0^{\varphi}(\vect u,\vect w)$ with respect to $\vect w$ given in Theorem~\ref{the:lip}.
\end{Theorem}
\begin{proof}
    The proof is given in the appendix.
\end{proof}
 
Note that the feasible domain of CCP~\eqref{eq:CCP} has been under-approximated by the tighter feasible domain of ECP~\eqref{eq:EP} with constraints on the expectation and tightening function $h^{-1}$. Problem~\eqref{eq:EP} can then be solved via: 1) computing the expectations $\mathbb E_P[\cdot]$, 2) utilizing knowledge of the family of distributions or an upper bound on the function $h$ related to the concentration of measure property, and 3) computing the robustness Lipschitz constant $L_\varphi$. 


In determining 2) and 3), knowledge of the exact distribution $P$ is not necessary. However, to compute expectations in 1) one requires $P$. For instance, sample averaging is an often employed technique to approximate expectations numerically. Nonetheless, this approach necessitates a sufficiently large sample size to ensure the accuracy of the empirical expectation compared to the exact one.





In the following, we aim to solve the ECP \eqref{eq:EP} for the unknown distribution $P$  with respect to the worst-case distribution in an ambiguity set using a Wasserstein distributionally robust approach and provide a finite sample guarantee with respect to the exact ECP~\eqref{eq:EP}. More specifically, the distributionally robust version of \eqref{eq:EP} can be written as the  following distributionally robust program~(DRP):
\begin{equation}
\label{eq:SP2}
\!\mathrm{DRP\!:}\!\begin{cases}
\min_{\vect u\in\mathbb U}\sup_{Q\in\mathbb{Q}} \mathbb E_{Q} \left[J(\vect u,\vect w)\right],\\
\mathrm{s.t.} \inf_{Q\in\mathbb{Q}}\mathbb E_Q \!\left[\varrho_0^{\varphi}(\vect u,\vect w)\right]\!-\!L_\varphi h^{-1}(\varepsilon)\!\geq\! r_0,
\end{cases}
\end{equation}
where $\mathbb{Q}$ is an ambiguity set, defining a set of all distributions around an empirical distribution $\hat Q$ that could contain the true distribution $P$ with high confidence. In this paper, we use the Wasserstein metric $W: Q(\mathbb{W})\times  Q(\mathbb{W}) \rightarrow \mathbb{R}_{\geq 0}$ to define the ambiguity ball $\mathbb{Q}$ as
\begin{equation}\label{eq:amb:set}
\mathbb{Q}:=\{Q\in Q(\mathbb{W})\,|\, W(Q,\hat Q)\leq r\},    
\end{equation}
where $Q(\mathbb{W})$ denotes the set of Borel probability measures on the support $\mathbb{W}$ and $r\geq 0$ is the radius of the Wasserstein ball. For any two distributions $Q^1,Q^2\in Q(\mathbb{W})$, the Wasserstein metric $W$ is defined as follows:
\begin{align}\label{eq:WM}
    &W(Q^1,Q^2):= \min_{\kappa\in Q(\mathbb{W}^2)} \Big\{\int_{\mathbb{W}^2} \|  \vect w_1- \vect w_2\|\mathrm{d}\kappa( \vect w_1, \vect w_2)\,\nonumber\\&\qquad\qquad\qquad\qquad\quad\qquad\Big|\, \Pi^j\kappa= Q^j, j=1,2 \Big\},
\end{align}
 where $\Pi^j\kappa$ denotes the $j^{\mathrm{th}}$ marginal of the joint distribution $\kappa$ for $j=1,2$. Note that, the sampling-based reformulation in~\eqref{eq:SP2} stems from the need to make decisions under uncertainty about the true distribution $P$ that governs the random variable $\vect w$. Since the true distribution $P$ is unknown, we rely on a finite number of i.i.d. samples $\{\vect w^i\}_{i=1}^M$ to infer information about $P$. These samples provide an empirical approximation $\hat{Q}$ that can be constructed as follows: 
\begin{equation}\label{eq:hatQ}
    \hat {Q}=\frac{1}{M} \sum_{i=1}^{M} \delta_{\vect w^i},
\end{equation}
where $\delta_{\vect w^i}$ is the Dirac measure concentrated at $\vect w^i$. Consider that, because $\hat{Q}$ is constructed from a limited sample set, it may not perfectly capture the true distribution $P$. To account for this uncertainty, we introduced the ambiguity set $\mathbb{Q}$ in~\eqref{eq:amb:set}, which includes all distributions that are close to the empirical distribution $\hat{Q}$ within a Wasserstein ball of radius $r$. The parameter $r$ reflects the confidence level: a larger $r$ increases the probability that $\mathbb{Q}$ contains the true distribution $P$, thereby providing a more robust solution to the optimization problem.

Although the DRP~\eqref{eq:SP2} overcomes knowledge of the exact distribution, it is tricky to solve in general since it contains decision variables in the continuous probability measure space. It is desirable to derive an (approximated) solution of~\eqref{eq:SP2} based on the finite samples $\vect w^i$, ensuring both feasibility and performance guarantees.

It is important to note that most of the date-driven DRO literature focuses on providing data-driven optimizations that are equivalent to the original problem in convex optimization~\cite{mohajerin2018data, chen2024data, xie2021distributionally}. In the field of stochastic optimization, many guarantees, such as those for approximating expectations from data, are also typically established when the function within the expectation is convex~\cite{shapiro2021lectures}.

However, since the robustness functions of STL are generally non-convex, it is necessary to develop an equivalent optimization approach that does not rely on this convexity assumption. Inspired by~\cite{mohajerin2018data} and results developed in~\cite{gao2023distributionally}, the following theorem offers a data-driven equivalent solution to the DRP~\eqref{eq:SP2} for the Wasserstein ambiguity set, defined in~\eqref{eq:amb:set}, and with a finite number of samples that eliminates the decision variables in the probability measure space. We can guarantee that any feasible solution obtained from the proposed optimization is a feasible solution to the DRP~\eqref{eq:SP2} and, with a predefined confidence level, is a feasible solution to the main ECP~\eqref{eq:EP}. Moreover, the obtained control input minimizes an upper bound on the objective in~\eqref{eq:EP} with high confidence.

\begin{Theorem}\label{the:data:DRO}\textbf{(Data-driven solution of the DRP)}\label{thm:1}
Consider the following optimization:
\begin{subequations}\label{eq:trac0}
\begin{flalign}
&\inf_{\vect u\in\mathbb U, \lambda_1,\lambda_2,y_1^i,y_2^i}\,\, \lambda_1 r+\frac{1}{M} \sum_{i=1}^{M} y_1^i,\\
      &\,\,\,\mathrm{s.t.}\,\, \sup_{\vect w\in \mathbb{W}} \left[J(\vect u, \vect w)-\lambda_1 \|  \vect w- \vect w^i\|\right]\leq y_1^i, \forall i\leq M,\label{eq:trac0:cons1}\\&\qquad\sup_{\vect w\in \mathbb{W}}\! \left[-\varrho_0^{\varphi}(\vect u,\vect w)\!-\!\lambda_2 \|  \vect w\!-\! \vect w^i\|\right]\!\leq\! y_2^i, \forall i\leq M,\label{eq:trac0:cons2}\\ &\qquad\lambda_1 \geq 0,\qquad \lambda_2 \geq 0, \\&\qquad
 -\lambda_2 r-\frac{1}{M} \sum_{i=1}^{M} y_2^i-L_\varphi h^{-1}(\varepsilon)\geq r_0,
\end{flalign} 
\end{subequations}
with the optimal solution and value of the objective function denoted by $\hat{\vect u}$ and $\hat J$, respectively. Based on the Wasserstein ambiguity set $\mathbb{Q}$ defined in \eqref{eq:amb:set}-\eqref{eq:hatQ}, we have:
\begin{itemize}
    \item \textbf{Relation to the DRP~\eqref{eq:SP2}:}
 Optimization \eqref{eq:trac0} is equivalent to \eqref{eq:SP2}.
 \item  \textbf{Relation to the ECP~\eqref{eq:EP}:}
By choosing a proper Wasserstein radius $r$, the following statements hold with the probability of at least $1-\beta$ for a user-specified confidence level $\beta\in(0, \,1)$:
\newline
1) Any feasible solution to~\eqref{eq:trac0} is a feasible solution to~\eqref{eq:EP}.
\newline
2) The cost function in~\eqref{eq:EP}, evaluated for the input $\hat{\vect u}$, is upper-bounded by $\hat J$. More specifically, the following out-of-sample performance guarantee holds:
\begin{equation}\label{eq:OOS}
    \mathbb{E}_{P}\left[J(\hat{\vect u}, \vect w) \right]\leq   \hat J.
\end{equation}
\end{itemize}
\end{Theorem}
\begin{proof}
    The proof is given in the appendix.
\end{proof}
Note that the term \textit{data-driven}, as used in e.g.,  \cite{mohajerin2018data, chen2024data, hota2019data}, refers to optimization that utilizes a set of samples to construct the DRP. In the control literature, this term has also been used in contexts where the system matrices are unknown. However, this is beyond the scope of our paper, which assumes that the system matrices are known in advance.

\begin{Remark}\label{rem:1}
The $\min$ and $\max$ operators utilized in defining robust semantic $\varrho_0^{\varphi}$ in Section~\ref{sec:stl} are not smooth. Numerical solvers commonly encounter difficulties when these operators appear in the objective function or constraints. Inspired by~\cite{gilpin2020smooth}, we opt for smooth under-approximations for these operators, as follows:
\begin{subequations}
\begin{align*}
\min ([a_1,\ldots,a_m]^\top) & \approx -\frac{1}{C} \log\left(\sum_{i=1}^m \exp({-Ca_i})\right),\\
\max ([a_1,\ldots,a_m]^\top) & \approx \frac{\sum_{i=1}^m a_i \exp({Ca_i})}{\sum_{i=1}^m \exp({Ca_i})},
\end{align*}
\end{subequations}
where $C$ is a positive constant. It is noteworthy that these approximations under-approximate the exact $\min$ and $\max$ operators. Consequently, the robust semantics derived from these approximations are not greater than the original robust semantics. Hence, fulfilling the approximated robust semantics ensures the satisfaction of the original semantics directly. Additionally, as demonstrated in~\cite{gilpin2020smooth}, for a sufficiently large $C$, the approximated robust semantics converge to the original semantics with the exact $\min$ and $\max$ operators.

Note that in transitioning from the CCP~\eqref{eq:CCP} to the ECP~\eqref{eq:EP}, we can assume that the robustness function $\varrho_0^{\varphi}$ is evaluated using the exact $\min$ and $\max$ operators, ensuring the validity of the Lipschitz constant $L_\varphi$ as obtained from Theorem~\ref{the:lip}. We then substitute the expectation of the exact robustness with the expectation of the under-approximated robustness in the constraint of~\eqref{eq:EP}. Therefore, the results of the paper, and particularly the Lipschitz constant $L_\varphi$, remain valid when employing these under-approximations of the exact $\min$ and $\max$ operators.

\end{Remark}

\section{Case study}\label{sec:examples}
We consider the following two-dimensional stochastic dynamics:
\begin{equation*}
    x_{k+1}=\begin{bmatrix}
1 & 1\\ 0 
 & 1
\end{bmatrix}x_{k}+\begin{bmatrix}
0.5\\ 1
\end{bmatrix}u_k+w_k,
\end{equation*}
where $u_k\in \mathcal{U}:=[-1,\, 1]$ and $w_k\in\mathcal{W}=\mathbb R^2$ with Gaussian distribution with unknown mean and covariance. Using \eqref{eq:state} and assuming $x_0=[-8\,,\, 0]^\top$, the aim is to satisfy a safety constraint $[0\,,\, 1]x_k\leq 0.75$ for the whole bounded horizon $0\leq k\leq N=15$ and reaching the region $  x^\top_kTx_k\leq 1$, with $T=\mathrm{diag}(\frac{1}{4},\frac{1}{25})$, sometime at $k\in[0,\,15]$ with probability at least $0.9$ while optimizing the following quadratic cost:
  \begin{equation*}
 J(\vect u,\vect w)= 10x^\top_{N}x_{N}+\sum_{k=0}^{N-1} (10x^\top_{k}x_{k}+u^2_{k}),
\end{equation*}  
where $x_k$ is obtained from~\eqref{eq:state}. The STL formula, described above, can be expressed as $\varphi= \F_{[0\, 15]} \pi_1 \land \G_{[0\, 15]} \pi_2$, where $\pi_1$ and $\pi_2$ are predicates with corresponding predicate functions $\alpha_1(x)=1-x^\top T x$ and $\alpha_2(x)=0.75-[0\,,\, 1]x$. The robustness function $\varrho_0^{\varphi}$ can be written as follows:
\begin{equation*}
    \varrho_0^{\varphi}(\vect u,\vect w)=\min\{\max_{k\in\{0,\ldots,15\}} \alpha_1(x_k),\min_{k\in\{0,\ldots,15\}} \alpha_2(x_k)\}.\nonumber 
\end{equation*}
As explained in Remark~\ref{rem:1}, we have chosen the smoothing constant $C$ as $100$, $10$, and $10$ for the inner minimization, the maximization, and the outer minimization in $\varrho_0^{\varphi}$, respectively.

\begin{figure}[t]
\centering
\includegraphics[width=0.48\textwidth]{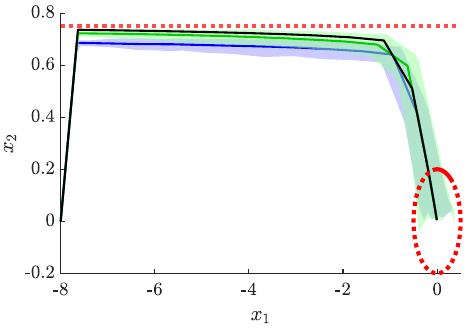}
\caption{System trajectories for different realizations for deterministic system (black), ECP solution using the sample average approximation (green), and the proposed DRP solution (blue).}
\label{fig:1}
\end{figure}

Figure~\ref{fig:1} shows the system trajectories using the proposed DRP approach for the Wasserstein radius $r=10^{-3}$. As it can be seen, the trajectories have greater distance with the bound $x_2=0.75$ compared to the sample averaging method and the STL specification is satisfied for all trajectories. We have employed $10$ times more sampling for the sample averaging method compared to the DRP method. 

\section{Conclusions}\label{sec:concl}

We have shown how to optimize control sequences for stochastic linear systems to satisfy signal temporal logic (STL) specifications probabilistically when the underlying predicate functions are Lipschitz continuous, and the disturbance distribution is unknown but attains a concentration of measure property. These assumptions allow us to reformulate the control problem as a chance-constrained program (CCP) and present an efficient two-step solution. First, leveraging the concentration of measure property, we transform the CCP into an expectation-based optimization problem. To account for unknown distributions, we proceed to the second step, where we tackle a distributionally robust optimization problem, which considers all distributions around the empirical one using an ambiguity set based on the Wasserstein metric. In the future, we plan to extend the method to multi-agent systems and nonlinear dynamics.









\bibliographystyle{IEEEtran}
\bibliography{DRP}  

\begin{thebibliography}{10}
\providecommand{\url}[1]{#1}
\csname url@samestyle\endcsname
\providecommand{\newblock}{\relax}
\providecommand{\bibinfo}[2]{#2}
\providecommand{\BIBentrySTDinterwordspacing}{\spaceskip=0pt\relax}
\providecommand{\BIBentryALTinterwordstretchfactor}{4}
\providecommand{\BIBentryALTinterwordspacing}{\spaceskip=\fontdimen2\font plus
\BIBentryALTinterwordstretchfactor\fontdimen3\font minus \fontdimen4\font\relax}
\providecommand{\BIBforeignlanguage}[2]{{%
\expandafter\ifx\csname l@#1\endcsname\relax
\typeout{** WARNING: IEEEtran.bst: No hyphenation pattern has been}%
\typeout{** loaded for the language `#1'. Using the pattern for}%
\typeout{** the default language instead.}%
\else
\language=\csname l@#1\endcsname
\fi
#2}}
\providecommand{\BIBdecl}{\relax}
\BIBdecl

\bibitem{Haesaert2018}
S.~Haesaert, P.~Nilsson, C.~Vasile, R.~Thakker, A.~Agha-mohammadi, A.~Ames, and R.~Murray, ``Temporal logic control of {POMDP}s via label-based stochastic simulation relations,'' \emph{IFAC-PapersOnLine}, vol.~51, no.~16, pp. 271--276, 2018.

\bibitem{donze2010robust}
A.~Donz{\'e} and O.~Maler, ``Robust satisfaction of temporal logic over real-valued signals,'' in \emph{International Conference on Formal Modeling and Analysis of Timed Systems}.\hskip 1em plus 0.5em minus 0.4em\relax Springer, 2010, pp. 92--106.

\bibitem{maler2004monitoring}
O.~Maler and D.~Nickovic, ``Monitoring temporal properties of continuous signals,'' in \emph{International Symposium on Formal Techniques in Real-Time and Fault-Tolerant Systems}.\hskip 1em plus 0.5em minus 0.4em\relax Springer, 2004, pp. 152--166.

\bibitem{farahani2017shrinking}
S.~S. Farahani, R.~Majumdar, V.~S. Prabhu, and S.~E.~Z. Soudjani, ``Shrinking horizon model predictive control with chance-constrained signal temporal logic specifications,'' in \emph{2017 American Control Conference (ACC)}.\hskip 1em plus 0.5em minus 0.4em\relax IEEE, 2017, pp. 1740--1746.

\bibitem{LindemannCDC2020}
L.~Lindemann, G.~J. Pappas, and D.~V. Dimarogonas, ``Control barrier functions for nonholonomic systems under risk signal temporal logic specifications,'' in \emph{Proceedings of the IEEE Conference on Decision and Control}, 2020, pp. 1422--1428.

\bibitem{Sadigh2016}
D.~Sadigh and A.~Kapoor, ``Safe control under uncertainty with probabilistic signal temporal logic,'' in \emph{Proceedings of Robotics: Science and Systems}, jun 2016.

\bibitem{Sadigh2018}
S.~Jha, V.~Raman, D.~Sadigh, and S.~A. Seshia, ``Safe autonomy under perception uncertainty using chance-constrained temporal logic,'' \emph{Journal of Automated Reasoning}, vol.~60, no.~1, pp. 43--62, 2018.

\bibitem{Li2017}
J.~Li, P.~Nuzzo, A.~Sangiovanni-Vincentelli, Y.~Xi, and D.~Li, ``{Stochastic contracts for cyber-physical system design under probabilistic requirements},'' in \emph{MEMOCODE 2017 - 15th ACM-IEEE International Conference on Formal Methods and Models for System Design}, 2017, pp. 5--14.

\bibitem{Scher2022}
G.~Scher, S.~Sadraddini, and H.~Kress-Gazit, ``Robustness-based synthesis for stochastic systems under signal temporal logic tasks,'' in \emph{2022 IEEE/RSJ International Conference on Intelligent Robots and Systems (IROS)}, 2022, pp. 1269--1275.

\bibitem{FarahaniTAC2019}
S.~S. Farahani, R.~Majumdar, V.~S. Prabhu, and S.~Soudjani, ``Shrinking horizon model predictive control with signal temporal logic constraints under stochastic disturbances,'' \emph{IEEE Transactions on Automatic Control}, vol.~64, no.~8, pp. 3324--3331, 2019.

\bibitem{Campi2011}
M.~C. Campi and S.~Garatti, ``{A Sampling-and-Discarding Approach to Chance-Constrained Optimization: Feasibility and Optimality},'' \emph{Journal of Optimization Theory and Applications}, vol. 148, no.~2, pp. 257--280, 2011.

\bibitem{Calafiore2010}
G.~C. Calafiore, ``Random convex programs,'' \emph{SIAM Journal on Optimization}, vol.~20, no.~6, pp. 3427--3464, 2010.

\bibitem{garatti2024non}
S.~Garatti and M.~C. Campi, ``Non-convex scenario optimization,'' \emph{Mathematical Programming}, pp. 1--52, 2024.

\bibitem{delage2010distributionally}
E.~Delage and Y.~Ye, ``Distributionally robust optimization under moment uncertainty with application to data-driven problems,'' \emph{Operations research}, vol.~58, no.~3, pp. 595--612, 2010.

\bibitem{hu2013kullback}
Z.~Hu and L.~J. Hong, ``Kullback-{L}eibler divergence constrained distributionally robust optimization,'' \emph{Available at Optimization Online}, pp. 1695--1724, 2013.

\bibitem{pflug2007ambiguity}
G.~Pflug and D.~Wozabal, ``Ambiguity in portfolio selection,'' \emph{Quantitative Finance}, vol.~7, no.~4, pp. 435--442, 2007.

\bibitem{mohajerin2018data}
P.~Mohajerin~Esfahani and D.~Kuhn, ``Data-driven distributionally robust optimization using the {W}asserstein metric: Performance guarantees and tractable reformulations,'' \emph{Mathematical Programming}, vol. 171, no.~1, pp. 115--166, 2018.

\bibitem{kordabad2022safe}
A.~B. Kordabad, R.~Wisniewski, and S.~Gros, ``Safe reinforcement learning using {W}asserstein distributionally robust {MPC} and chance constraint,'' \emph{IEEE Access}, vol.~10, pp. 130\,058--130\,067, 2022.

\bibitem{chen2024data}
Z.~Chen, D.~Kuhn, and W.~Wiesemann, ``Data-driven chance constrained programs over {W}asserstein balls,'' \emph{Operations Research}, vol.~72, no.~1, pp. 410--424, 2024.

\bibitem{xie2021distributionally}
W.~Xie, ``On distributionally robust chance constrained programs with {W}asserstein distance,'' \emph{Mathematical Programming}, vol. 186, no.~1, pp. 115--155, 2021.

\bibitem{hota2019data}
A.~R. Hota, A.~Cherukuri, and J.~Lygeros, ``Data-driven chance constrained optimization under {W}asserstein ambiguity sets,'' in \emph{2019 American Control Conference (ACC)}.\hskip 1em plus 0.5em minus 0.4em\relax IEEE, 2019, pp. 1501--1506.

\bibitem{Soudjani2018}
S.~Soudjani and R.~Majumdar, ``Concentration of measure for chance-constrained optimization,'' \emph{IFAC-PapersOnLine}, vol.~51, no.~16, pp. 277--282, 2018, 6th IFAC Conference on Analysis and Design of Hybrid Systems ADHS 2018.

\bibitem{MalNic:04}
O.~Maler and D.~Nickovic, ``Monitoring temporal properties of continuous signals,'' in \emph{{FORMATS/FTRTFT}}, ser. LNCS 3253.\hskip 1em plus 0.5em minus 0.4em\relax Springer, 2004.

\bibitem{req_mining_hscc2013}
X.~Jin, A.~Donz{\'{e}}, J.~V. Deshmukh, and S.~A. Seshia, ``Mining requirements from closed-loop control models,'' \emph{{IEEE} Trans. on {CAD} of Integrated Circuits and Systems}, vol.~34, no.~11, pp. 1704--1717, 2015.

\bibitem{barvinok1997measure}
A.~Barvinok, ``Measure concentration in optimization,'' \emph{Mathematical Programming}, vol.~79, pp. 33--53, 1997.

\bibitem{pisier1999volume}
G.~Pisier, \emph{The volume of convex bodies and Banach space geometry}.\hskip 1em plus 0.5em minus 0.4em\relax Cambridge University Press, 1999, vol.~94.

\bibitem{shapiro2021lectures}
A.~Shapiro, D.~Dentcheva, and A.~Ruszczynski, \emph{Lectures on stochastic programming: modeling and theory}.\hskip 1em plus 0.5em minus 0.4em\relax SIAM, 2021.

\bibitem{gao2023distributionally}
R.~Gao and A.~Kleywegt, ``Distributionally robust stochastic optimization with {W}asserstein distance,'' \emph{Mathematics of Operations Research}, vol.~48, no.~2, pp. 603--655, 2023.

\bibitem{gilpin2020smooth}
Y.~Gilpin, V.~Kurtz, and H.~Lin, ``A smooth robustness measure of signal temporal logic for symbolic control,'' \emph{IEEE Control Systems Letters}, vol.~5, no.~1, pp. 241--246, 2020.

\bibitem{fournier2015rate}
N.~Fournier and A.~Guillin, ``On the rate of convergence in {W}asserstein distance of the empirical measure,'' \emph{Probability Theory and Related Fields}, vol. 162, no.~3, pp. 707--738, 2015.

\end{thebibliography}
\newpage
\section*{Appendix}

   \begin{Lemma}\label{lem:ab}
Suppose that $a_1, a_2, b_1, b_2\in\mathbb R$ are such that $a_1\leq a_2$ and $b_1\leq b_2$. Then:   
\begin{equation*}
        |a_1-b_1|\leq \max\{|a_1-b_2|,|a_2-b_1|\}.
\end{equation*}
\end{Lemma} 
\begin{proof}
    Suppose that $a_1\leq b_1$. Then $a_1\leq b_1\leq b_2$ and $|a_1-b_1|\leq |a_1-b_2|$. In the case that  $b_1\leq a_1$, we have $b_1\leq a_1\leq a_2$ and $|a_1-b_1|\leq |a_2-b_1|$.
\end{proof}

\subsection{Proof of Lemma~\ref{lem:lip:minmax}}

\begin{proof}
   Since the functions $f_1$ and $f_2$ are Lipschitz continuous, we have:
    \begin{equation*}
         |f_i(x_1)-f_i(x_2)|\leq L_id_X(x_1,x_2), \,\, i\in\{1,2\}, \,\, \forall x_1,x_2\in X     
    \end{equation*}
    for some $ L_1, L_2\geq 0$, where $d_X$ is a metric on the set $X$. For any $x_1, x_2\in X$, we have:
\begin{align}\label{ineq:lem2}
    &|\min(f_1(x_1), f_2(x_1))-\min(f_1(x_2),f_2(x_2))|\leq \max \Big\{
   \nonumber \\ &|f_1(x_1)-f_1(x_2)|, |f_1(x_1)-f_2(x_2)|, |f_2(x_1)-f_1(x_2)|, \nonumber\\  &|f_2(x_1)-f_2(x_2)|\Big\}\leq \max \Big\{|f_1(x_1)-f_1(x_2)|,\nonumber \\ &|f_2(x_1)-f_2(x_2)|\Big\}\leq \max\{L_1,L_2\}d_X(x_1,x_2),
\end{align}
where the first inequality follows from different cases that might happen for the two $\min$ operators, and the second inequality can be concluded from Lemma~\ref{lem:ab} for $|f_1(x_1)-f_2(x_2)|$ and $|f_2(x_1)-f_1(x_2)|$ cases. For instance, for $|f_1(x_1)-f_2(x_2)|$ case, we can set $a_1=f_1(x_1)$, $a_2=f_2(x_1)$, $b_1=f_1(x_2)$ and $b_2=f_2(x_2)$. From the facts that $\min(f_1(x_1), f_2(x_1))=f_1(x_1)$ and $\min(f_1(x_2), f_2(x_2))=f_2(x_2)$, the inequalities of Lemma~\ref{lem:ab}, stating $a_1\leq a_2$ and $b_1\leq b_2$ are fulfilled and therefore 
\begin{align*}
 |f_1(x_1)-f_2(x_2)|\leq \max \Big\{&|f_1(x_1)-f_1(x_2)|,\nonumber\\ &|f_2(x_1)-f_2(x_2)|\Big\}.   
\end{align*}
A similar statement holds for the case of $|f_2(x_1)-f_1(x_2)|$ in the second inequality of \eqref{ineq:lem2}. Similarly, we can show the lemma holds for the $\max$ operator.
\end{proof}

\subsection{Proof of Theorem~\ref{the:lip}}
\begin{proof}
By Assumption~\ref{Assum:predicate}, Lemma~\ref{lem:lip:minmax} and the system trajectories in~\eqref{eq:state}, the mapping $\vect w\rightarrow \varrho_0^{\varphi}$ is Lipschitz continuous for all $\vect u$.  
To obtain the Lipschitz constant, it can be shown that for the composition of functions, the Lipschitz constant is the multiplication of Lipschitz constants of all functions. 
Therefore, using Lemma~\ref{lem:lip:minmax} and its extension to multiple $\min$ and $\max$ operators, it is straightforward to show that the robustness function is a Lipschitz function with constant $L_{\varphi}=L_1L_2$, where $L_1$ is the maximum Lipschitz constant of atomic predicates appearing in $\varphi$, and $L_2$ is the maximum Lipschitz constant of system trajectory with respect to $\vect w$ among all times in $1\leq k\leq N$. To find the Lipschitz constant of system trajectory with respect to $\vect w$ for a given time $k$ and for the linear system given in~\eqref{eq:discsys}, we use~\eqref{eq:state} with setting $x_0=0$ and $u_i=0$ for all $0\leq i\leq k-1$. Then we have $x_{k} = \sum_{i=0}^{k-1}A^{k-i-1} w_i$, with the following upper bound on the norm:

\begin{align*}
\|x_k\| & =\left\|\sum_{i=0}^{k-1}A^{k-i-1} w_i\right\|\leq \sum_{i=0}^{k-1}\|A^{k-i-1} w_i\|\\
& \leq\!
\sum_{i=0}^{k-1}\!\|A^{k-i-1}\| \|w_i\|\leq\!
\sqrt{\sum_{i=0}^{k-1}\!\|A^{k-i-1}\|^2}
\sqrt{\sum_{i=0}^{k-1}\|w_i\|^2} \\
&\le  \|\vect w\| \sqrt{\sum_{i=0}^{k-1}\|A^i\|^2}\le L_2 \|\vect w\|.
\end{align*}
Note that we have used the Cauchy–Schwarz inequality and the obtained bound is valid for all $1\leq k\leq N$. 
\end{proof}

\subsection{Proof of Theorem~\ref{the:EP}}
   
\begin{proof}
From Theorem~\ref{the:lip}, the function $\varrho_0^{\varphi}(\vect u,\vect w)/L_\varphi$ is Lipschitz continuous with respect to $\vect w$ with Lipschitz constant $1$. Under Assumption~\ref{ass:concentration} with $t = h^{-1}(\varepsilon)$:
\begin{equation*}\label{eq:concen:phi}	 P\left\{\left|\frac{\varrho_0^{\varphi}(\vect u,\vect w)}{L_\varphi}-\mathbb E_P\left[\frac{\varrho_0^{\varphi}(\vect u,\vect w)}{L_\varphi}\right]\right |\le t\right\}\ge 1-h(t)=1-\varepsilon,
	\end{equation*}
for all $t\ge 0$ and $\vect u\in\mathbb U$. Suppose that the constraint in the ECP~\eqref{eq:EP} holds, i.e., 
\begin{equation*}
    \mathbb E_{P} \left[\frac{\varrho_0^{\varphi}(\vect u,\vect w)}{L_\varphi}\right]\geq h^{-1}(\varepsilon)+ \frac{r_0}{L_\varphi}=t+\frac{r_0}{L_\varphi}.
\end{equation*}
Then, with probability at least $(1-\varepsilon)$, we have:  
\begin{align*}
\frac{\varrho_0^{\varphi}(\vect u,\vect w)}{L_\varphi}\ge \mathbb E_P\left[\frac{\varrho_0^{\varphi}(\vect u,\vect w)}{L_\varphi}\right]-t\ge t+\frac{r_0}{L_\varphi} -t = \frac{r_0}{L_\varphi}.
\end{align*}
This implies that $\varrho_0^{\varphi}(\vect u,\vect w)\ge r_0$ with probability at least $(1-\varepsilon)$.
\end{proof}
\subsection{Proof of Theorem~\ref{the:data:DRO}}\label{ap:a}
\begin{proof}
First, we investigate the relation between~\eqref{eq:trac0} and~\eqref{eq:SP2}. Using the definition of Wasserstein metric in~\eqref{eq:WM}, $ \sup_{Q\in\mathbb{Q}}\mathbb E_Q \left[J(\vect u,\vect w)\right]$ can be reformulated as 
\begin{align}\label{eq:opt:DR1}
   \sup_{Q\in\mathbb{Q}}\mathbb E_Q &\left[J(\vect u,\vect w)\right]=
   \begin{cases}
       \sup_{Q,\kappa} \mathbb E_Q \left[J(\vect u,\vect w)\right],\\
       \mathrm{s.t.} \int_{\mathbb{W}^2} \|  \vect w_1- \vect w_2\|\mathrm{d}\kappa( \vect w_1, \vect w_2)\leq r\,,\nonumber\\\qquad \Pi^1\kappa= Q,\,\, \Pi^2\kappa= \hat Q
   \end{cases}
   \\&=
     \begin{cases}
       \sup_{Q_i} \frac{1}{M} \sum_{i=1}^{M} \mathbb E_{Q_i} \left[J(\vect u,\vect w)\right]\\
       \mathrm{s.t.} \frac{1}{M} \sum_{i=1}^{M} \mathbb E_{Q_i} \left[\|  \vect w- \vect w^i\|\right]\leq r.
   \end{cases}
\end{align}
The second equality is a consequence of the law of total probability, which states that any joint distribution $\kappa$ of $\vect w_1$ and $\vect w_2$ can be created from the marginal distribution $\hat Q$ of $\vect w_2$ and the conditional distribution $Q_i$ of $\vect w_1$ given $\vect w_2=\vect w^i$, i.e., $\kappa= \frac{1}{M} \sum_{i=1}^{M} \delta_{ \vect w^i} \otimes Q_i$, where $\otimes$ is used for the product of two probability distributions.
Using the Lagrangian dual problem for the constrained optimization \eqref{eq:opt:DR1}, we obtain:
\begin{align}\label{eq:Lag}
    &\sup_{Q\in\mathbb{Q}}\mathbb E_Q \left[J(\vect u,\vect w)\right]=\sup_{Q_i}\inf_{\lambda_1\geq 0} \frac{1}{M} \sum_{i=1}^{M} \mathbb E_{Q_i} \left[J(\vect u,\vect w)\right]+\nonumber\\ &\qquad\qquad\lambda_1 \left(r-\frac{1}{M} \sum_{i=1}^{M} \mathbb E_{Q_i} \left[\|  \vect w- \vect w^i\|\right]\right)=\nonumber\\
& \inf_{\lambda_1\geq 0} \sup_{Q_i}\lambda_1 r+ \frac{1}{M} \sum_{i=1}^{M} \mathbb E_{Q_i} \left[J(\vect u, \vect w)-\lambda_1\|  \vect w- \vect w^i\|\right]= \nonumber\\&
\inf_{\lambda_1\geq 0}\lambda_1 r+\frac{1}{M} \sum_{i=1}^{M} \sup_{\vect w\in \mathbb{W}} \left[J(\vect u, \vect w)-\lambda_1\|  \vect w- \vect w^i\|\right],  
\end{align}
where $\lambda_1$ is the Lagrange multiplier. The first equality in \eqref{eq:Lag}, once bounded, follows from the strong duality that has been shown in \cite{gao2023distributionally}, and the second equality holds because $Q_i$ contains all the Dirac distributions supported on $\mathbb W$, i.e., one can verify that $\sup_X \mathbb{E}^X[f(x)]=\sup_x f(x)$, where $X$ is any distribution on the support of random variable $x$. Introducing a new auxiliary variable $y_1^i$, from \eqref{eq:Lag} we can conclude:
\begin{align}\label{eq:supinf1}
&\sup_{Q\in\mathbb{Q}}\mathbb E_Q \left[J(\vect u,\vect w)\right]= \\ &
\begin{cases}\nonumber
    \inf_{\lambda_1, y_1} &\lambda_1 r+\frac{1}{M} \sum_{i=1}^{M} y_1^i,\\
    \mathrm{s.t.}  &\sup_{\vect w\in \mathbb{W}}\! \left[J(\vect u, \vect w)\!-\!\lambda_1 \|  \vect w- \vect w^i\|\right]\!\!\leq\! y_1^i, \forall i\leq M,\\ &\lambda_1 \geq 0.
\end{cases}
\end{align}
Similarly, we can show the following:
\begin{align}\label{eq:supinf2}
&\inf_{Q\in\mathbb{Q}}\mathbb E_Q \left[\varrho_0^{\varphi}(\vect u,\vect w)\right]=-\sup_{Q\in\mathbb{Q}}\mathbb E_Q \left[-\varrho_0^{\varphi}(\vect u,\vect w)\right]= \\ &
\begin{cases}\nonumber
    \sup_{\lambda_2,y_2^i}\quad -\lambda_2 r-\frac{1}{M} \sum_{i=1}^{M} y_2^i,\\
    \mathrm{s.t.} \,\,\, \sup_{\vect w\in \mathbb{W}} \left[-\varrho_0^{\varphi}(\vect u,\vect w)-\lambda_2 \|  \vect w- \vect w^i\|\right]\leq y_2^i, \forall i\leq M,\\ \qquad\lambda_2 \geq 0.
\end{cases}
\end{align}
We can utilize the equalities in~\eqref{eq:supinf1} and ~\eqref{eq:supinf2} in the main DRP~\eqref{eq:SP2} to obtain the data-driven optimization in~\eqref{eq:trac0}. Therefore, the feasible domain of the data-driven optimization in~\eqref{eq:trac0} is equal to DRP~\eqref{eq:SP2} and the objective functions are the same, i.e.,:


\begin{equation}\label{eq:hatJ1}
   \sup_{Q\in\mathbb{Q}} \mathbb E_{Q} \left[J(\hat {\vect u},\vect w)\right]= \hat J.
\end{equation}

Regarding the connection to the ECP~\eqref{eq:EP} with the exact distribution, \cite{fournier2015rate} has demonstrated that if the Wasserstein radius $r$ is selected as follows:
\begin{equation}\label{eq:epsi}
    r\geq \left \{ \begin{matrix}
\left(\frac{\log(c_1\beta^{-1})}{c_2M}\right)^{\frac{1}{\max\{Ns,2\}}} & \mathrm{if}\quad M\geq \frac{\log(c_1\beta^{-1})}{c_2},
\\
\left(\frac{\log(c_1\beta^{-1})}{c_2M}\right)^{\frac{1}{a}} & \mathrm{else},
\end{matrix} \right.
\end{equation}
then the probability that the Wasserstein ball $\mathbb Q$ contains the exact distribution is at least $1-\beta$, i.e., $
    {P}^M\left \{P\in \mathbb Q\right \}\geq 1-\beta,$ where $c_1, c_2$ are positive constants depending on $a$ and $C$ (see Assumption~\ref{ass:light-tailed}) and the disturbance dimension~\cite{fournier2015rate}. Therefore, for every given $\vect u$:
    \begin{equation*}
    {P}^M\left \{\inf_{Q\in\mathbb{Q}}\mathbb E_Q \left[\varrho_0^{\varphi}(\vect u,\vect w)\right]\leq \mathbb E_P \left[\varrho_0^{\varphi}(\vect u,\vect w)\right]\right\}\geq 1-\beta,
\end{equation*}
where ${P}^M$ is the $M$-fold product of the probability measure $P$. This concludes that the feasible domain of \eqref{eq:SP2} (and \eqref{eq:trac0} from the first part of the theorem) is an inner approximation of~\eqref{eq:EP} with a probability of at least $1-\beta$. Similarly for the objective function and for all $\vect u\in\mathbb U$, including $\hat{\vect u}$, we have:
\begin{equation}\label{eq:PM}
       {P}^M\left \{\sup_{Q\in\mathbb{Q}}\mathbb E_Q \left[J(\hat{\vect u},\vect w)\right]\geq \mathbb E_P \left[J(\hat{\vect u},\vect w)\right]\right\}\geq 1-\beta.
\end{equation}
From~\eqref{eq:hatJ1} and \eqref{eq:PM}, \eqref{eq:OOS} can be obtained, and this concludes the proof.
\end{proof}

\end{document}